\RequirePackage{fix-cm}
\documentclass[smallextended]{svjour3}       
\smartqed  
\usepackage{soul}
\usepackage{bm,textcomp}

\usepackage{graphicx}
\usepackage{color}
\usepackage{natbib}
\journalname{Journal of Membrane Biology}
\newcommand{\bemb}{n_{\mathrm{emb}}}
\newcommand{\bann}{n_{\mathrm{ann}}}
\newcommand{\femb}{f_{\mathrm{emb}}}
\newcommand{\fann}{f_{\mathrm{ann}}}

\begin{document}

\title{Untangling direct and domain-mediated interactions between nicotinic acetylcholine receptors in DHA-rich membranes}


\author{Kristen Woods*         \and
        Liam Sharp* \and 
        Grace Brannigan$^\dagger$ 
}


\institute{Kristen Woods and Liam Sharp and Grace Brannigan \at
              Center for Computational and Integrative Biology, Rutgers University-Camden, Camden, NJ \\
           \and
         Grace Brannigan \at
             Department of Physics, Rutgers University-Camden, Camden, NJ
             \and
             *These authors contributed equally.  
            $^\dagger$Corresponding Author (grace.brannigan@rutgers.edu) 
}

\date{Received: date / Accepted: date}
\titlerunning{Interactions among nAChRs in DHA-rich membranes}

\maketitle
\begin{abstract}

 At the neuromuscular junction (NMJ), the nicotinic acetylcholine receptor (nAChR) self-associates to give rise to rapid muscle movement. While lipid domains have maintained nAChR aggregates in-vitro, their specific roles in nAChR clustering are currently unknown. In the present study, we carried out coarse-grained molecular dynamics simulations (CG-MD) of 1-4 nAChR molecules in two membrane environments: One mixture containing domain-forming, homoacidic lipids, and a second mixture consisting of heteroacidic lipids. Spontaneous dimerization of nAChRs was up to ten times more likely in domain-forming membranes; however, the effect was not significant in four-protein systems, suggesting that lipid domains are less critical to nAChR oligomerization when protein concentration is higher. With regard to lipid preferences, nAChRs consistently partitioned into liquid-disordered domains occupied by the omega-3 ($\omega$-3) fatty acid, Docosahexaenoic acid (DHA); enrichment of DHA boundary lipids increased with protein concentration, particularly in homoacidic membranes. This result suggests dimer formation blocks access of saturated chains and cholesterol, but not polyunsaturated chains, to boundary lipid sites.   


\end{abstract}

\section{Introduction}
\label{intro}
The muscle-derived nicotinic acetylcholine receptor (nAChR) (PDB 2BG9) \citep{Unwin2005} is the most abundant neurotransmitter receptor at the neuromuscular junction (NMJ) in most vertebrates, including humans \citep{Albuquerque2009}. Within the postsynaptic membrane, nAChRs cluster in high densities (10,000 per $\mu$m$^{2}$) to properly activate the skeletal muscle \citep{Ramarao1998,Breckenridge1972}. As a major transmembrane protein, nAChR depends upon a highly specific lipid environment to maintain functionality. Lipids influence nAChR activity by affecting both function and organization. It is essential to understand how changes in lipid environment impact nAChR's structure and activity, given that lipids can change in response to aging and disease  \citep{Yadav2014}, and also vary across tissue and organism.

Over the past few decades, considerable progress has been made in uncovering lipid sensitivities associated with nAChR \citep{Criado1982}. In a majority of experiments, researchers have prioritized studying cholesterol over other membrane lipids. Early studies revealed that, when reconstituted into membrane mixtures, nAChR failed to conduct cations across the lipid bilayer unless cholesterol was present  \citep{Fong_Correlation_1986,Sunshine_Lipid_1992,Butler1993,Fong_Stabilization_1987,Corrie_Lipid_2002}. 
More recently, researchers have examined the effects of membrane dynamics on nAChR organization \citep{Baenziger2015,Bruses2001,Marchand2002a,Oshikawa2003,Pato2008,Zhu2006a,Baenziger2017,Barrantes2007,Barrantes2000,Bermudez2010,Barrantes2010,Perillo2016,Wenz2005,Borroni2016,Unwin2017}. In-vitro studies \citep{Barrantes2007,Barrantes2010} indicate that nAChRs form larger aggregates upon cholesterol depletion. Experimental evidence suggest that cholesterol-rich lipid domains, known as lipid rafts, facilitate clustering of nAChRs \citep{Campagna2006,Marchand2002a,Pato2008}. 
More specifically, after disrupting lipid raft formation, Zhu et al. observed a significant loss of nAChR clusters in-vitro \citep{Zhu2006a}. In the mature neuromuscular membrane, nAChRs are linked by the intracellular anchoring protein, rapsyn, which bridges receptors together at their bases  \citep{Zuber2013a}. According to fluorescent studies \citep{Marchand2002a}, lipid rafts mediate the association between rapsyn and neighboring nAChR molecules. In the mid-2000's, Willmann et al. and Stetzkowski-Marden et al. proposed that lipid rafts can stabilize receptor networks and may even provide a localized environment for nAChR \citep{Willmann2006,Stetzkowski-Marden2006}. 

Domain formation occurs when a membrane is comprised of at least three lipid types: cholesterol, unsaturated fatty acids, and a molecule that interacts closely with cholesterol such as saturated fatty acids or sphingomyelin \citep{Feller_Acyl_2008,Yeagle2016115}.  The membrane separates into at least two domains, a liquid-ordered or ``raft'' phase containing cholesterol and saturated lipids/sphingomyelin, and a liquid-disordered phase containing unsaturated lipids. Polyunsaturated phospholipids make these domains more well-defined \citep{Levental_Polyunsaturated_2016}. Some results from atomistic simulations \citep{Sodt2014, Iyer2018} indicate that cholesterol may preferentially partition to the boundary between liquid-ordered phases and liquid-disordered phases composed of monounsaturated lipids.  In neuromuscular membranes, intrinsic domain formation is dependent upon several lipid species, including the widely influential omega-3 ($\omega$-3) fatty acids. One $\omega$-3 in particular, Docosahexaenoic acid (DHA), is prevalent in the native neuromuscular membrane and is strongly associated with flexible and well-defined domains  \citep{Turk2013,shaikh_dumaual_castillo_locascio_siddiqui_stillwell_wassall_2004}. Additionally, DHA is a major contributor to brain functioning, motor activity, and cardiac health; however, its specific effects on neuromuscular health are poorly understood \citep{12439486320170901,S000930840800032720080101,Georgieva2015}. 

\begin{figure}[htp]
\centering
\includegraphics[width=.99\columnwidth,trim={0cm 2cm 0cm 0cm}]{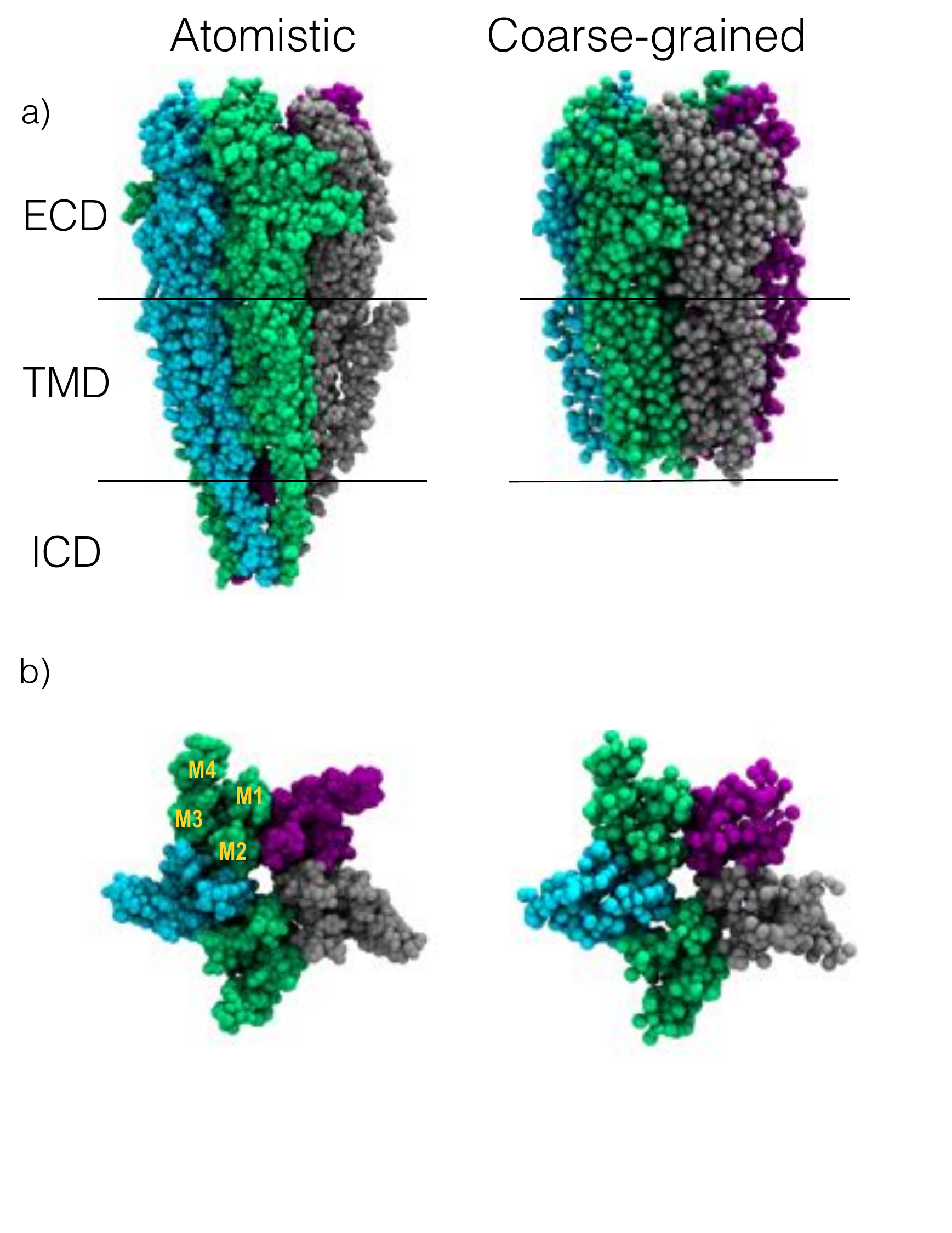}
\caption[Nicotinic acetylcholine receptor (nAChR) structure]{{\bf Nicotinic acetylcholine receptor (nAChR) structure.} a) Atomistic and coarse-grained representations of Unwin et al's cryo-EM structure, PDB 2BG9. The nAChR is colored by subunit ($\alpha:green,\beta:purple,\delta:gray,\gamma:cyan$) and labeled by structure. The extracellular domain (ECD) is located above the bilayer and is critical for ligand-binding. The transmembrane domain (TMD) is positioned within the lipid bilayer, and the intracellular domain (ICD) is located in the cytoplasm. The coarse-grained model omits the ICD since it is poorly resolved and not necessary for this study. b) TMD from the extracellular perspective, with its four alpha helices labelled in a single $\alpha$ subunit. The outermost M4 helices closely interact with surrounding membrane lipids, while the M2 helices outline the ion pore; the M1 and M3 helices make up the body of the transmembrane domain \citep{Unwin2005}.}
  \label{fig:structure}
\end{figure}

Characterizing such complex lipid-protein interactions requires detailed information on each molecular structure. The nAChR is a member of a family of ion channels, known as pentameric ligand-gated ion channels (pLGICs), which has a number of recent structures \citep{Laverty2019,Laverty2017,Masiulis2019,Althoff2014,Hibbs2011,Morales2016,Baenziger2011,Corringer2012,S089662731630023X20160504,Prevost2012a,Sauguet2014}; however, only single structures have been determined, rather than dimers. pLGICs are composed of five subunits; each subunit is composed of an extracellular domain for ligand binding, a transmembrane domain (TMD), and an intracellular domain (ICD) (Figure 1A). The TMD is the embedded portion of pLGICs that interacts most often with surrounding membrane lipids; this region of pLGICs is composed of four alpha-helices, M1-M4, with the innermost, M2 helices forming the ion pore. The ICD, or cytoplasmic domain, is highly disordered, making it challenging to obtain information on its structure. Currently, there are two structures of nAChR pentamers resolved from x-ray crystallography (neuronal type) and cryo-electron microscopy (muscle-type) \citep{osti1330893, Unwin2005}.

Other closely related channels have been resolved with bound lipids. In 2017, the inhibitory neurotransmitter receptor, the gamma-aminobutyric acid (GABA$_A$) receptor, was crystallized with a cholesterol molecule protruding between its M1 and M3 helices \citep{Laverty2017}, similar to our previous prediction \citep{Henin2014}. Interestingly, x-ray structures of the glutamate-gated chloride channel (GluCl) revealed embedded POPC phospholipids in the same location \citep{Althoff2014}. Together, these data provide evidence for lipid-based modulation of pLGICs, which can potentially be applied to the gating of nAChR.

Structural information alone, however, is insufficient for answering questions about native cell membranes. For one, most structures are obtained under artificial conditions, using detergents or nanodisks with non-native lipid composition. Additionally, current structures do not represent oligomers, particularly in a liquid state. Fully atomistic molecular dynamics (MD) simulations \citep{brannigan, Cheng2009, Hnin_A_2014, Carswell_Role_2015} have complemented experimental investigations of pLGICs interacting with lipids, but they are limited in their ability to capture domain formation since atomistic lipids cannot diffuse over typical simulation microsecond time scales \citep{Ingolfsson2014,Bond2006,Parton2013,Goose2013,Scott2008}.  Furthermore, it is unfeasible to incorporate multiple nAChRs in simulations with atomistic resolution. Coarse-grained MD (CG-MD) simulations are run over longer length and time scales, making them suitable for exploring complex model membranes. CG-MD is widely applied in simulations of both lipid-protein binding and domain formation \citep{Bond2006,Scott2008,Parton2013,Goose2013}. Additionally, CG-MD can capture large-scale membrane phenomenon such as protein self-assembly and lipid-mediated oligomerization \citep{10.3389/fphys.2016.00494, BAADEN2013878}.

Recently, we \citep{Sharp2019} conducted CG-MD simulations of a single nAChR from the electric ray {\it Torpedo} in mixed membranes. Contrary to expectations, nAChR consistently preferred a local lipid environment rich in PUFAs rather than cholesterol, especially long-chained $\omega$-3s, such as DHA. While cholesterol occupied the transmembrane gaps of nAChR, PUFAs were even more likely to be embedded, regardless of their phospholipid headgroup. 

The present study adopts a similar approach, with a particular focus on nAChR-associated clustering. Through molecular dynamics simulations, we investigate nAChR lipid preferences and clustering behavior in membranes with and without domains. For this study, we tested three major hypotheses: 1) Membrane organization affects nAChR boundary lipid specificity: when PUFA chains are prevented from forming PUFA rich domains, their prevalence among nAChR boundary lipids will be significantly reduced. 2) Domain formation will indirectly facilitate the clustering of nAChRs, by inducing partitioning preferences and restricting diffusion within the membrane. 3) Within a dimer, we will observe sequence preferences in facing subunits.
\section{Methods}
\subsection{System setup}

In order to isolate the role of lipid domain formation on nAChR-based lipid sorting and clustering, our simulations compared two different membranes with distinct phospholipid topology. Each membrane had 30\% cholesterol and 70\% phospholipids.  The phospholipids all had phosphatidylcholine (PC) headgroups, and half the total number of acyl chains were saturated (16:0) chains, while the other half were polyunsaturated (22:6) chains.  For homoacid membranes, all phospholipids had either two polyunsaturated acyl chains (didocosahexaenoylphosphatidylcholine or dDHA-PC) or two saturated chains (dipalmitoylphosphatidylcholine or DPPC), facilitating domain-formation. In the heteroacid membranes, all phospholipids were hybrid lipids with one polyunsaturated acyl chain and one saturated chain (1-palmitoyl- 2-docosahexaenoyl- phosphatidylcholine or PDPC), which are topologically unable to separate into PUFA-rich and PUFA-poor domains. 

Lipids and proteins were modeled using the coarse-grained (CG) Martini 2.2 force field \citep{martini}. 
Systems included between 1-4 nAChR molecules, derived from the Torpedo electric organ \citep{Unwin2005} (PDB 2BG9).  This structure is the only pLGIC structure obtained in a native membrane. 

We converted protein structures into CG models using the Martini script "martinize.py", mapping four non-hydrogen atoms to one CG interaction. We constructed and assembled our protein-bilayer systems using  the Martini script, "insane.py", using a box sizes of 29x29x21 nm$^{3}$ for one-to-two proteins, 40x40x20 nm$^{3}$ for three proteins, and 44x44x22 nm$^{3}$ for four protein systems, respectively  \citep{Marrink2007}. Initially, the proteins were in a circle of about 13 nm. The receptors were evenly spaced, and their $\delta$ subunits were facing the same direction. Once simulations started, nAChRs shifted from their initial orientations. We ran 24 CG-MD simulations containing 1-4 nAChRs (3 replicas per system). 

\subsection{Simulation details}
Simulations were run using the Martini 2.2 force field parameters and the Gromacs 5.1.2 simulation package, \citep{Marrink2007,Pronk2013} as in our previous work, \citep{Sharp2019}. Each simulation consisted of two steps: energy minimization and molecular dynamics. {For each system, we ran two consecutive energy minimizations for 10,000 steps. }
Harmonic restraints between backbone atoms were imposed to preserve nAChR conformation. 
More specifically, we applied an elastic force constant of 750 kJ/mol and set lower and upper bounds on the bond with using a bond length of 0.7 nm
 \citep{Marrink2007,Pronk2013}. The molecular dynamics simulations ran for 10-20 $\mu$s at a 0.025 ps time-step. Simulation temperature and pressure were kept constant at values of 323 K and a reference pressure of 1 bar. The isotropic pressure coupling compressibility constant was maintained at 3.0 $\times {10^{-5}}$ bar$^{-1}$. 



\subsection{Analysis}

For a given nAChR, $\bemb^{\alpha}$ is the total number of embedded lipids of lipid species $\alpha$, where embedded lipids satisfy the following criteria: the headgroup (in the case of cholesterol) or the terminal bead on the acyl chains (in the case of phospholipids) are within 10 {\AA}~of the M2 helices. Visual inspection indicated any lipids outside of this range were also outside of the TMD bundle.

Similarly, $\bann^{\alpha}$ is the total number of annular lipids of lipid species $\alpha$, where annular lipids satisfy the following criteria:  headgroup or the terminal bead on the acyl chains is between 10 {\AA}~and 35 {\AA}~of the M2 helices. This range of lipids corresponded to the third lipidation shell around nAChR.

(For phospholipids, each acyl chain was counted separately as a half-lipid, so it was possible for e.g. the sn-1 chain to be embedded and the sn-2 chain to be annular. )

Boundary lipid fractions for a given species $\alpha$ are defined as 
\begin{eqnarray}
      \femb^{\alpha}&\equiv \frac{  \bemb^{\alpha} }{\bemb },\\
      \fann^{\alpha}&\equiv \frac{\bann^{\alpha}}{\bann}
    \label{eq:f}
  \end{eqnarray}
where $\bemb$ and $\bann$ are the total number of embedded and annular lipids, respectively.  

Two dimensional density distributions of boundary acyl chain species (B), $\tilde{\rho}_{B}(r_i,\theta_j)$ were calculated, as a function of radius, $r_i$, and angle $\theta_j$ projected onto the membrane.

  \begin{equation}
      \rho_{B}(r_i,\theta_j)= \frac{\left\langle n_{B}(r_i,\theta_j) \right\rangle}{r_i \Delta{r}\Delta{\theta}} \\        
    \label{eq:R}
  \end{equation}

where $r_i = i \Delta r$ is the distance from the origin to the center of bin i, $\Delta r$ is the radial bin width, and $\Delta \theta$ is the angular bin width. $\left\langle n_{B}(r_i,\theta_j) \right\rangle$ is the time-averaged number of acyl chain beads of species B within bin i.

To quantify enrichment or depletion of a given acyl chain with respect to random distribution, the normalized density, $ \tilde{\rho}_{B}(r_i,\theta_j)$, was calculated.

  \begin{equation}
  \tilde{\rho}_{B}(r_i,\theta_j)=\frac{ \rho_{B}(r_i,\theta_j)}{x_{B}s_{B}~N_{L}/\langle L^{2}\rangle} \\        
    \label{eq:Rt}
  \end{equation}
  
  where $s_B$ is the number of beads of lipid species B, $N_L$ is total number of lipids in a system, and $\langle L^{2}\rangle$ is the average projected box area, and $x_B$ is lipid B concentration. The expression does not take into consideration the protein footprint or undulations present within the system, and as such is an approximation.

The radial distribution function $g(r)$ of multiple nAChRs was calculated using the three dimensional distance $r$ between the centers of mass of the receptors for each of the 8000 frames, evenly distributed through the simulation. 
It was derived from the distribution of pairwise separations, $P(r)$, by dividing by the expected separations in a random distribution : $g(r) = \frac{P(r)~dr}{r~dr}\times\int_{0}^{R} r dr$.  
The height of the $g(r)$ peak corresponds to the amount of enrichment in probability for a given distance; for example, if g(10~nm) = 50, it is 50 times more probable that two monomers will be 10~nm apart than expected in a random distribution.

The total number of observed dimers $n_{d}$ is given by the number of pairs (across all analyzed proteins) where $r<100$ {\AA}. For each observed dimer, the closest subunit pair was determined, and the enrichment calculated as $F_{x,y}$:

\begin{equation}\label{eq:subunit}
F_{x,y} = \frac{{25}}{n_{d}}\left({n_{x,y}}+ n_{y,x}\right),
\end{equation}
where $n_{x,y}$ represents the number of dimers in which subunits $x$ and $y$ form the closest pair, and $1/25$ is the expectation for a random distribution.

We visualized and imaged all simulation results using Visual Molecular Dynamics (VMD) \citep{HUMP96}. 
\section{Results}
\begin{figure}[h]
\center
\includegraphics[width=\linewidth,trim={0cm 0cm 0cm 0cm}]{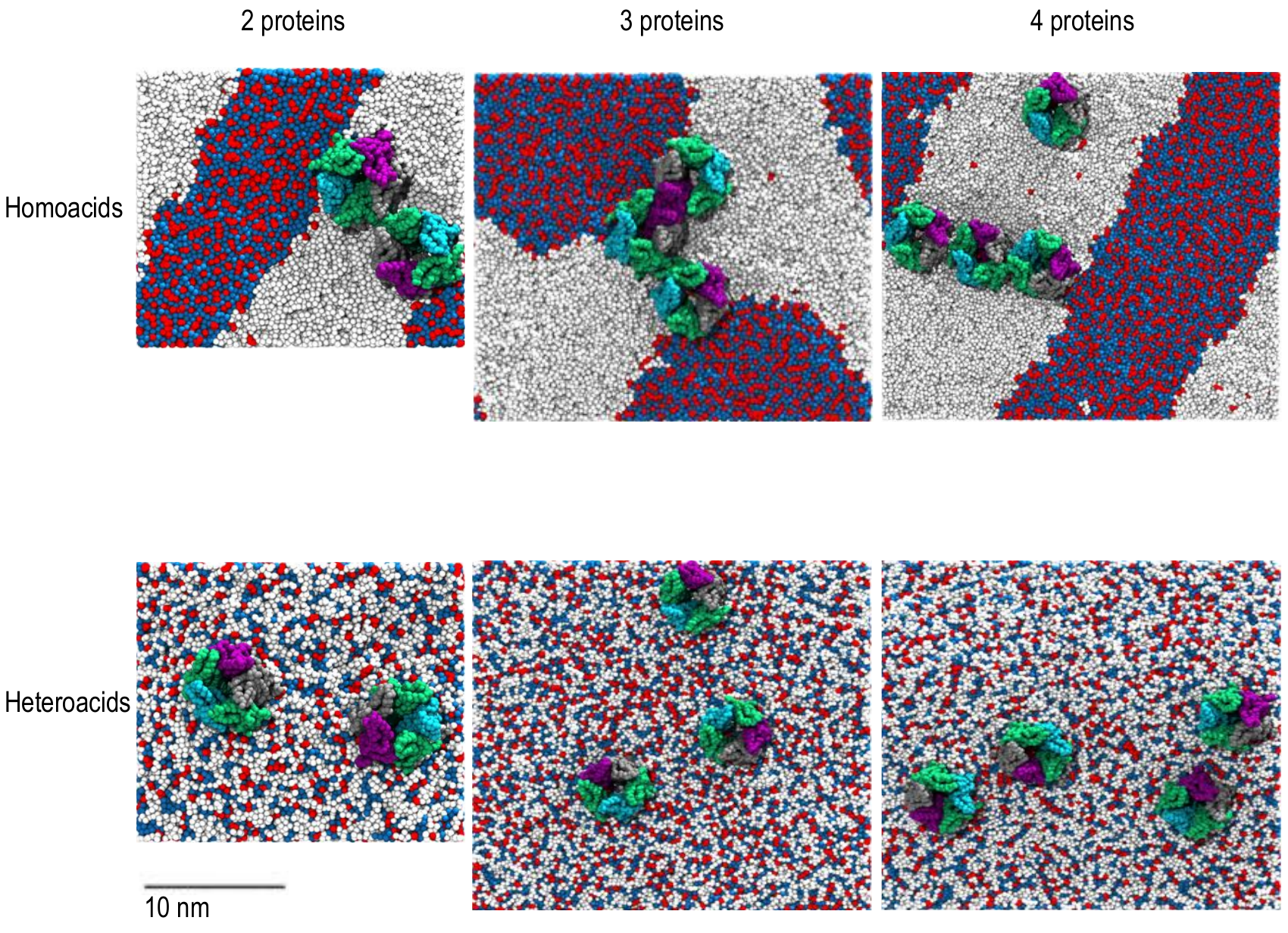}
\caption[Protein clustering in domain-forming (top row) and hybrid (bottom row) membranes.]{{\bf Protein clustering in domain-forming (top row) and hybrid (bottom row) membranes.} View of the membrane from the extracellular region at the final frame of 10 $\mu s$ simulations. nAChRs were colored by subunit ($\alpha:green,\beta:purple,\delta:gray,\gamma:cyan$), and lipids by acyl chain (DHA: white, saturated: blue, and cholesterol: red). Clustering of nAChRs on the borders of lipid rafts is visible in the domain-forming membranes.}
\label{fig:Figure4}
\end{figure}

\subsection{nAChR boundary lipid preferences}\label{sec:lipids}

\begin{table}[htbp]
  \centering
  \begin{tabular}{@{} c|cc|cc|c @{}}
     &\multicolumn{2}{c|}{mean embedded fraction}  &\multicolumn{2}{c|}{mean annular fraction} &bulk  \\ 
     & homoacid & heteroacid & homoacid & heteroacid  & \\ 
    \hline
    cholesterol & 0.29 (0.02)  & 0.42 (0.04) &0.23 (0.02)& 0.31 (0.01)& 0.30  \\ 
    PUFA  &0.61 (0.01)& 0.39 (0.04) & 0.52 (0.02) & 0.32 (0.01)&0.35\\ 
    saturated & 0.10 (0.05) & 0.19 (0.01) & 0.25 (0.02) & 0.37  (0.01)&0.35 \\ 
    \hline
  \end{tabular}
  \caption{Composition of nAChR Boundary Lipid chains in domain forming (homoacidic) and non-domain forming (heteroacidic) membranes.  Embedded and Annular chains are determined as described in Methods.  Averages are across systems, and across proteins in multi-protein systems.  Each protein is treated as a separate replica (n=30) and standard errors are shown in parentheses.  }
  \label{tab:bound}
\end{table}
     
%
%
%
%

In order to determine the significance of domain formation on enrichment of polyunsaturated acyl chains among boundary lipids, we calculated the fraction of embedded and annular boundary lipids in two types of membranes with equal amounts of cholesterol and saturated and polyunsaturated acyl chains.  In the domain-forming membranes, all phospholipids had either two polyunsaturated acyl chains or two saturated chains (homoacids), while in the non-domain forming membranes, all phospholipids had one polyunsaturated acyl chain and one saturated chain (heteroacids).  

\begin{figure}[htp]
\centering
\includegraphics[width=\textwidth]{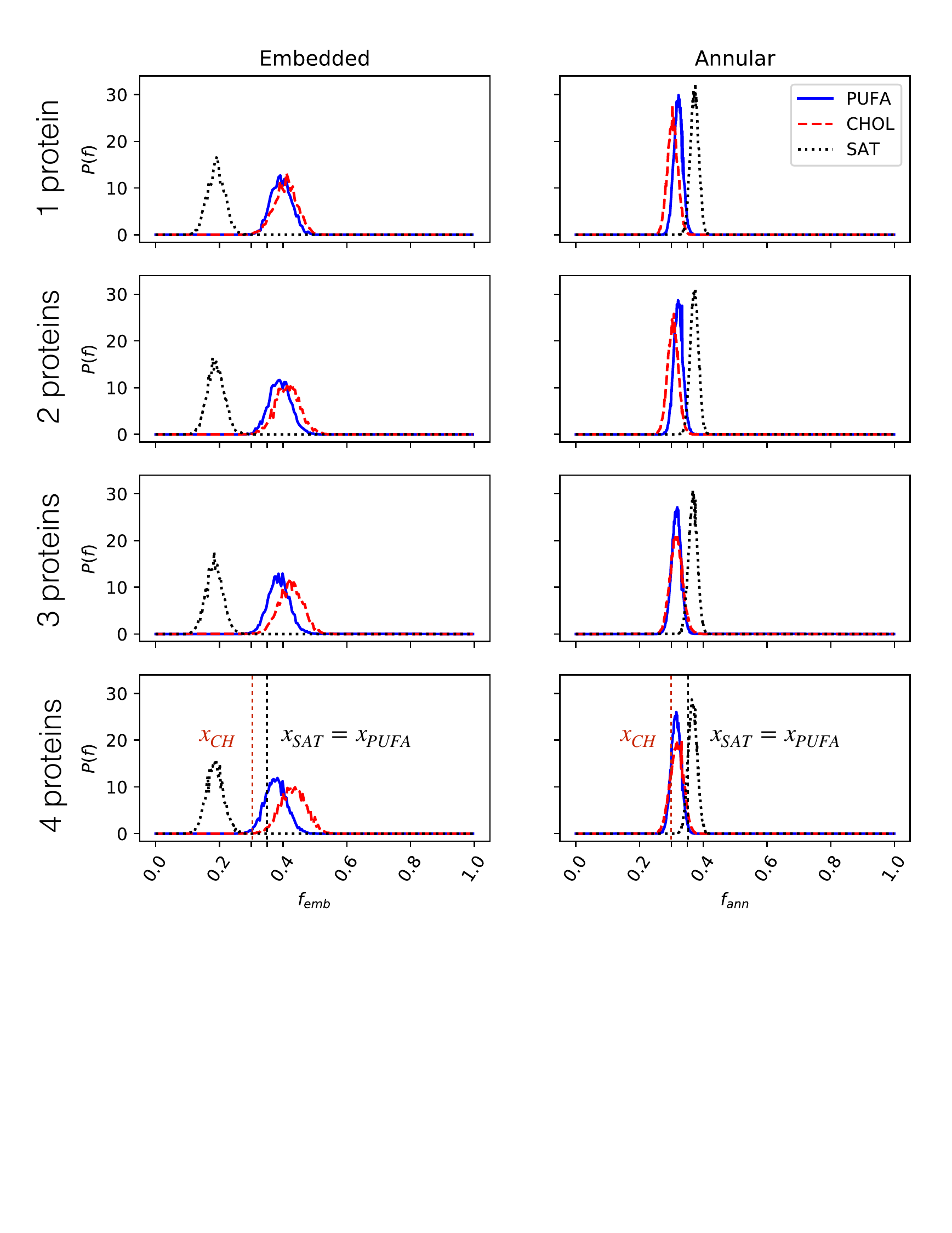}
\caption[nAChR boundary lipid preferences.]{{\bf Distribution of nAChR boundary cholesterol or acyl chain fractions in mixed membranes containing heteroacidic phospholipids.} Probability density distribution of fraction of embedded ($f_{emb}$) or annular lipids ($f_{ann}$) as defined in Eq. \ref{eq:f} are shown for 1 to 4 proteins.  Dashed lines represent expected boundary ratios for a randomly-mixed membrane, based on bulk lipid composition ($x_{CH} = 0.3$, $x_{SAT}=x_{PUFA}= 0.35$)}
  \label{fig:Fig2a}
\end{figure}

\begin{figure}[htp]
\centering
\includegraphics[width=\textwidth]{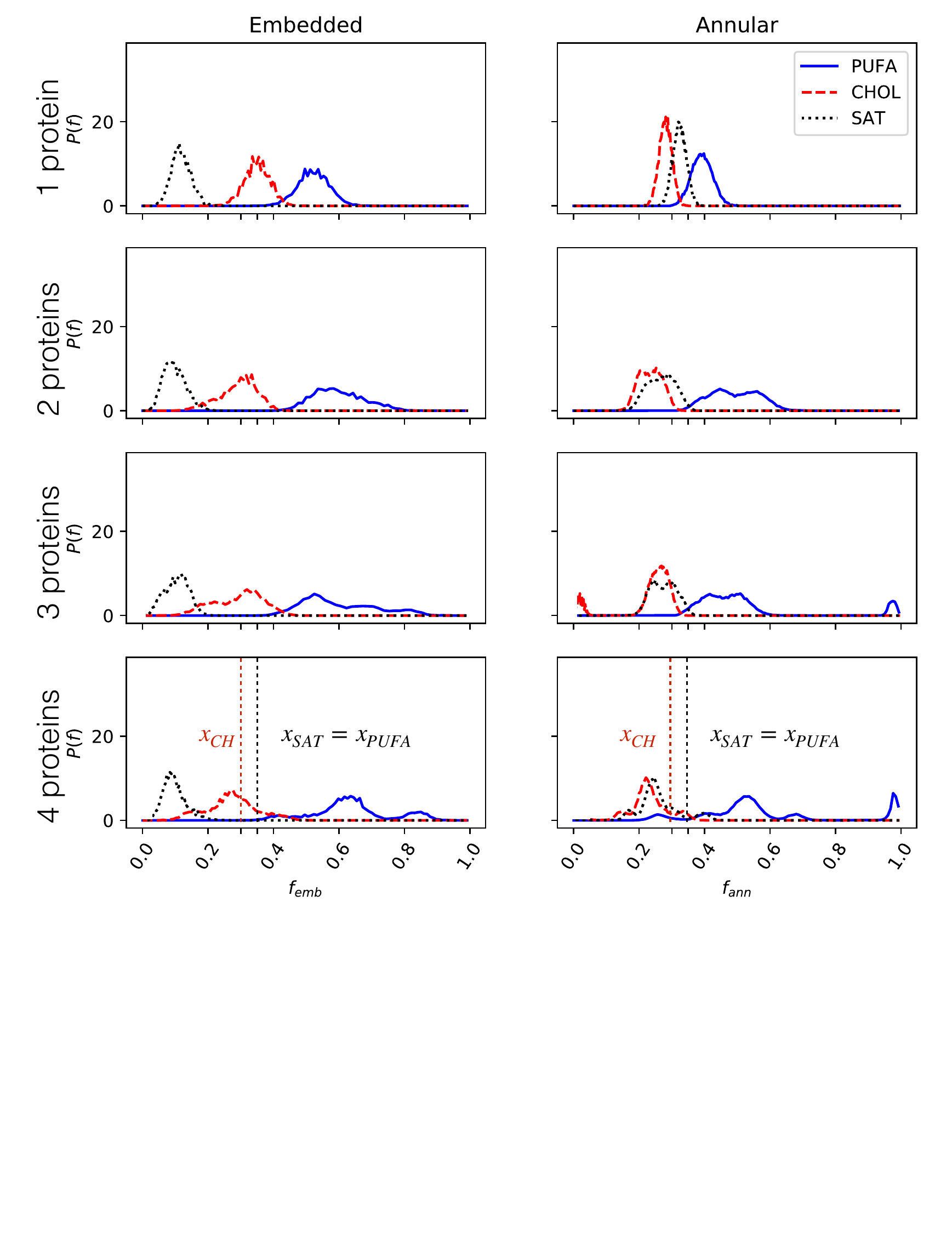}
\caption[nAChR boundary lipid preferences.]{{\bf Distribution of nAChR boundary cholesterol or acyl chain fractions in mixed membranes containing homoacidic phospholipids.} Probability density distribution of fraction of embedded ($f_{emb}$) or annular lipids ($f_{ann}$) as defined in Eq. \ref{eq:f} are shown for 1 to 4 proteins.  Dashed lines represent expected boundary ratios for a randomly-mixed membrane, based on bulk lipid composition ($x_{CH} = 0.3$, $x_{SAT}=x_{PUFA}= 0.35$).} \label{fig:Fig2b}
\end{figure}

Distributions for $f_{emb}$ and $f_{ann}$ are shown in Figure \ref{fig:Fig2a} in heteroacids compositions without domains and in Figure \ref{fig:Fig2b} for homoacids compositions with domains, while mean values are given in Table \ref{tab:bound}.  For heteroacids, PUFA chains are slightly enriched among embedded chains and slightly depleted among annular chains, while the reverse trend is observed for saturated lipids.  Saturated chains are significantly depleted from the embedded lipids, but due to the topology of these lipids, a fully embedded polyunsaturated acyl chain must also contribute a saturated chain to the protein annulus. Distributions were remarkably consistent across simulations containing 1,2,3, or 4 proteins.


Figure \ref{fig:Fig2b} shows the results of the same analysis for homoacidic membranes.  Replacing heteroacids with homoacids substantially increases the fraction of polyunsaturated chains among both embedded and annular lipids, with the peak of the distribution shifting further to the right as more proteins are included (Figure \ref{fig:Fig2b}). 

Together, these results are consistent with our previous \citep{Sharp2019} observation that polyunsaturated chains can displace cholesterol from embedded sites.  For heteroacids,  embedding a PUFA constrained the linked saturated chain to the nAChR annulus. In this case, cholesterol (which introduces no such constraint) was enriched among embedded lipids.  For homoacids, embedded PUFAs constrained a corresponding PUFA chain to the nAChR annulus, and here cholesterol was not enriched among embedded lipids. 

The fraction of embedded chains that were polyunsaturated increased as more proteins were added to the homoacidic membrane. This could be consistent with each nAChR monomer in an oligomer blocking access to embedded lipid binding sites in the other monomers: flexible PUFA chains have multiple routes to access an embedded site, and rigid lipids have only one or two.  

\begin{figure}[htp]
\centering
\includegraphics[width=\textwidth]{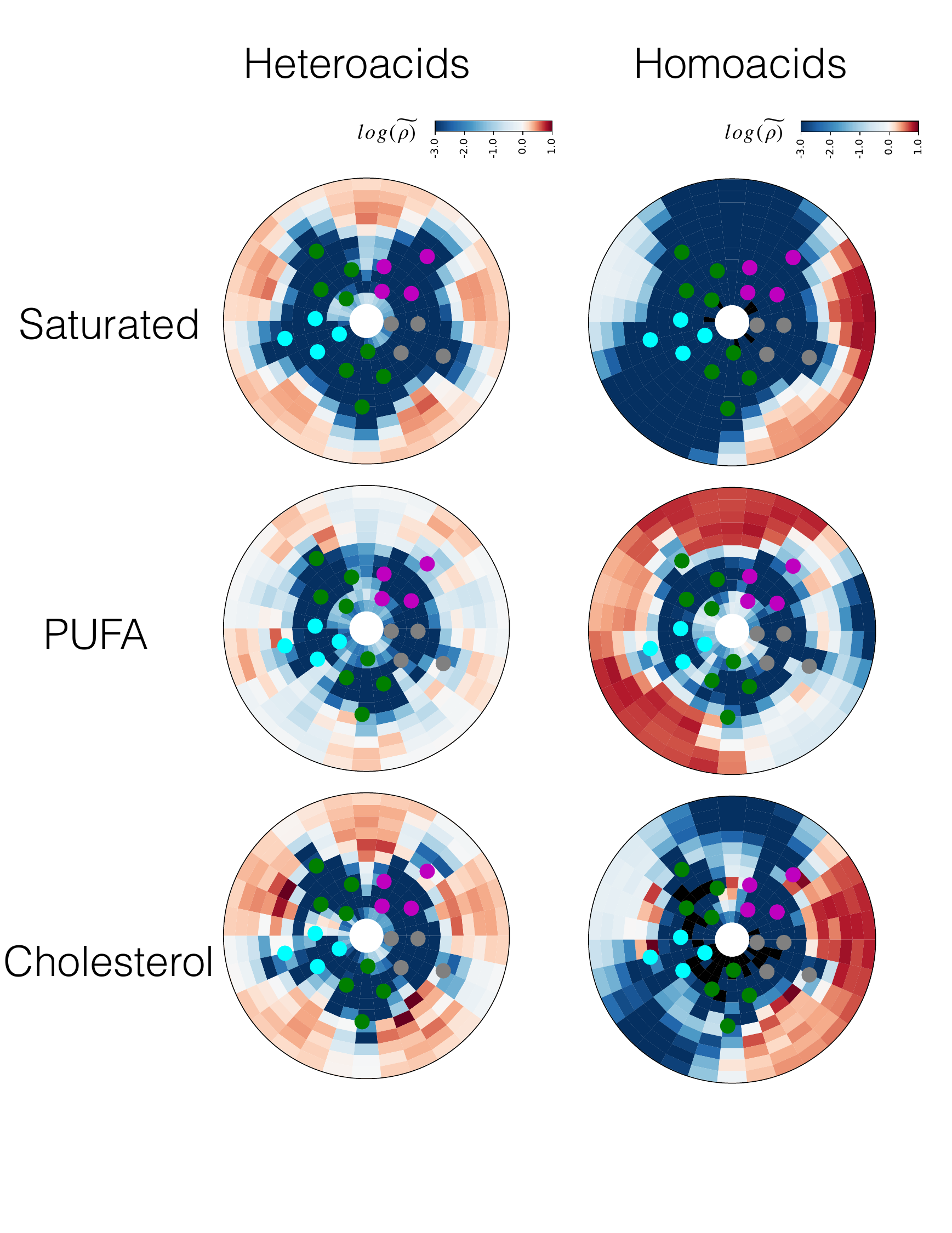}
\caption[Polar plot.]{{\bf Enrichment or depletion of lipid species surrounding a single nAChR.} Heatmaps represent normalized lipid densities ($\tilde{\rho}_{B}(r_i,\theta_j)$) as defined in Eq. \ref{eq:Rt}, on a scale of -3 (blue, indicating depletion) to 1 (red, indicating enrichment). Densities were averaged over the last 8 $\mu s$ of the 10 $\mu s$ simulations and across three replicas.} \label{fig:Fig2c}
\end{figure}

The two dimensional density distribution of each lipid species from the center of a single nAChR is shown in Figure \ref{fig:Fig2c}. For all three lipids, there was a five-fold symmetry of densities surrounding nAChR, with lipid preferences determined by transmembrane helix, rather than by subunit. In heteroacid mixtures, cholesterol is enriched near the subunit interfaces, and even buried more deeply within the $\gamma/\alpha$ interface, with saturated acyl chains packed just outside cholesterol.  The outermost M4 helices, however, were packed with PUFAs; PUFA chains also diffused throughout the TMD bundle.  In homoacid membranes, saturated lipids and cholesterol were depleted among embedded lipids, but they maintained the same helix association observed in heteroacid membranes. PUFAs were especially enriched in domain-forming membranes, with high densities around all four transmembrane helices. 

\subsection{nAChR clustering in the presence and absence of lipid domains} \label{sec:clustering}
\begin{figure}[htp]
\centering
\includegraphics[width=1.0\linewidth]{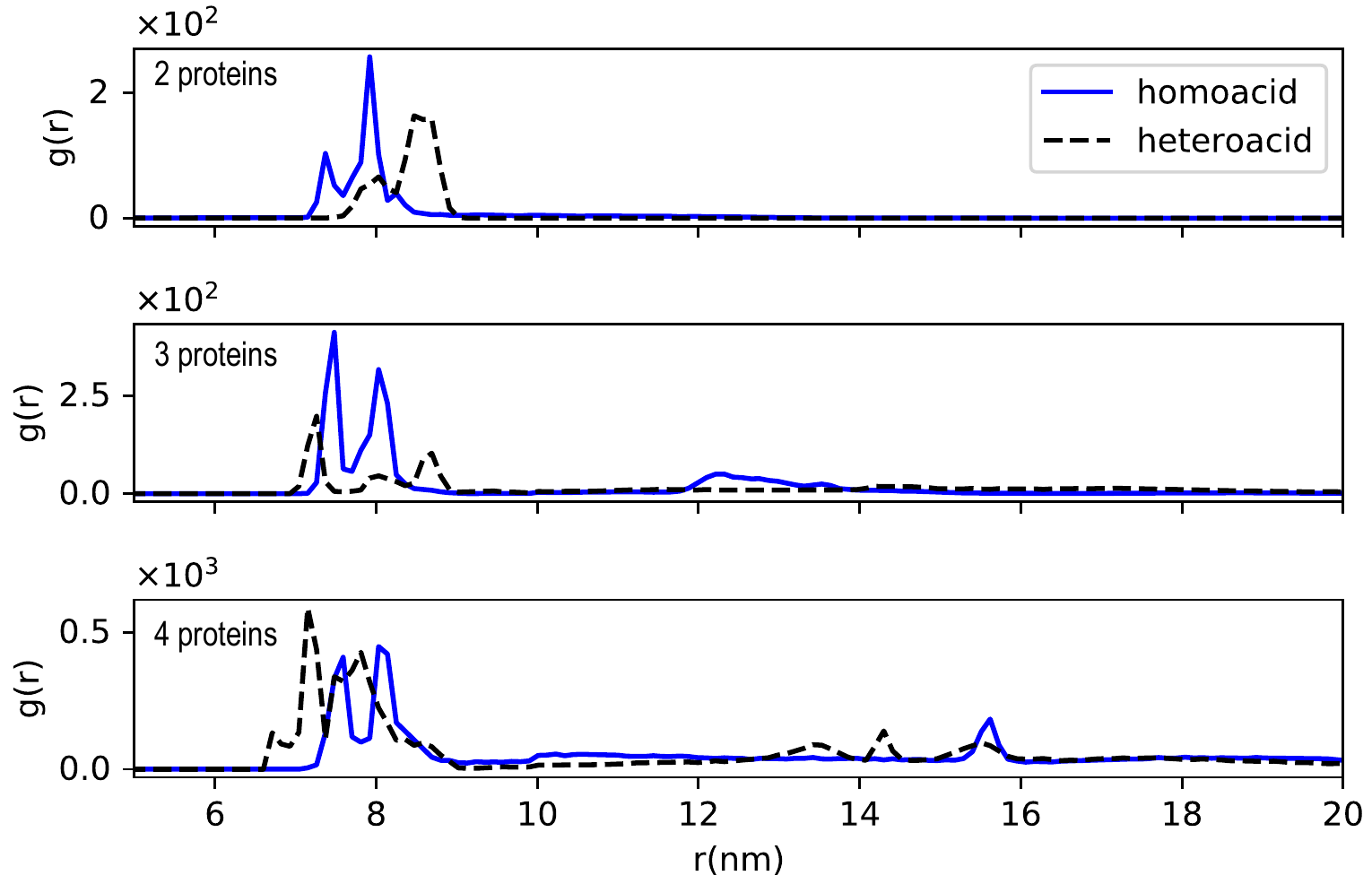}
\caption[Pairwise distance distribution across multi-protein systems.] {{\bf Radial distribution function g(r)  of pairwise protein center-of-mass distances, across multi-protein systems.}  Data collection began at 2 $\mu s$ into each trajectory and ended at 10 $\mu s$, respectively. The peak between 7-10 nm corresponds to dimer formation.  }
\label{fig:Figure6}
\end{figure}

Here, we repeatedly observed spontaneous formation of receptor dimers in the simulations containing multiple proteins (Figure \ref{fig:Figure4}). 
To investigate the role of lipid domain formation in driving nAChR oligomerization, we calculated the radial distribution function for the pairwise distances between centers of mass, as shown in Figure \ref{fig:Figure6}. For both heteroacid and homoacid membranes, we observed a peak at $r\sim7.5$~nm, corresponding to dimerization.    The differences in profiles between domain and non-domain forming membranes were primarily quantitative: most significantly, the peak corresponding to dimers was substantially higher for 2 and 3 proteins in homoacids than in heteroacids. For two proteins, the distribution for homoacids was shifted to the left, relative to the distribution for heteroacids, indicating that the peak for dimerization was at a shorter distance in domain-forming membranes. This difference is consistent with results indicating a role for domain formation in aggregation and clustering of nAChRs \citep{Barrantes2007,Oshikawa2003,Pato2008}. The simplest explanation of this difference is the higher effective concentration of proteins when domains are present: all nAChRs are corralled in a single liquid-disordered domain, with approximately half the area of the overall membrane.  

\subsection{Closest subunits across dimerizing proteins}
\begin{figure}[htp]
	\center
	\includegraphics[height=3in]{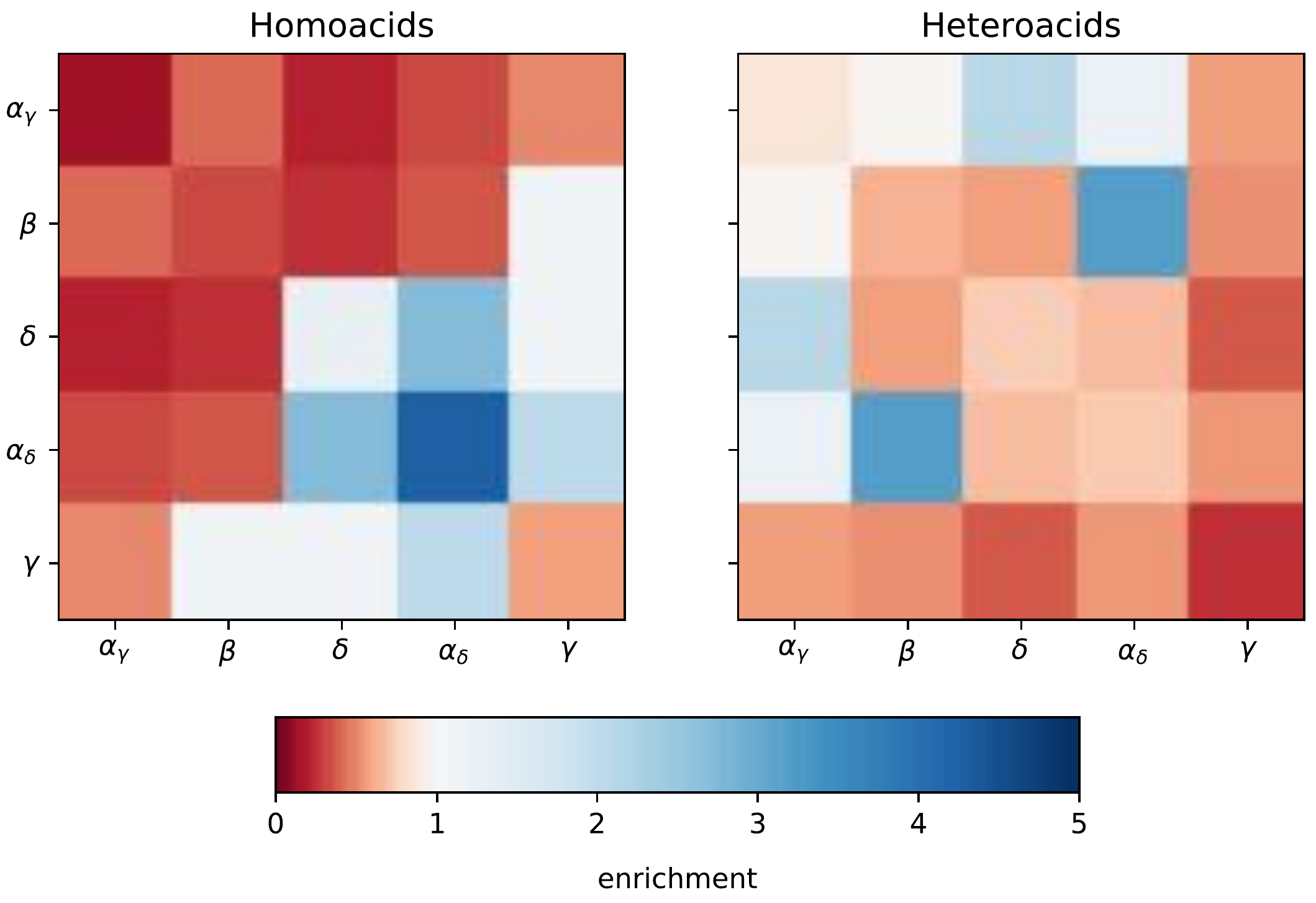}
\caption[Subunit pairs among dimerizing proteins]{{\bf Most probable subunit pairs among dimerizing proteins.} Heat-map showing the likelihood that two subunits were closest together among dimerizing proteins (100 {\AA} threshold). Distinct subunits are each expected to form the closest pair 8 \% of the time, while identical subunits are each expected to dimerize only 4 \% of the time. Each color on the heatmap represents multiplicative depletion or enrichment relative to the pair's expectation value, ranging from complete depletion (red) to no enrichment (white) to five-fold enrichment (blue).  }
\label{fig:Figure7}
\end{figure}

In order to determine whether nAChR dimers were more likely to form with specific subunit interactions, we determined the closest interacting subunit pair for nAChR dimers within each frame. To ensure that dimers, rather than larger oligomers, were being analyzed, we excluded trimers and tetramers from the analysis and only considered two protein systems, with increased sampling. Figure ~\ref{fig:Figure7} shows the amount of enrichment for each possible subunit pairing, relative to an expected random distribution. Results were very sensitive to the use of domain-forming compositions. In heteroacidic membranes, the $\alpha_{\delta}$ subunit formed a monomer-monomer interaction with the $\beta$ subunit, while the $\alpha_{\gamma}$ most favorably interacted with the $\delta$ subunit. In domain-forming membranes,  the $\delta-\alpha_{\delta}-\gamma$ interface was two to five times as likely to pair with a $\alpha_{\delta}$ subunit compared to a random distribution, but substantially more dispersion was detected than in non-domain forming membranes. This difference suggests that nAChRs have preferred dimerization orientations, but membrane organization can reorient receptors. Although $\delta$ subunits are linked by a disulfide bond at the NMJ \citep{Chang1977}, our simulations suggest that without this link, $\delta$ subunits do not form the closest pair among dimerizing proteins.


\section{Discussion} 


The nAChR is one of the most well-studied, fundamental pLGICs for understanding human cognition, memory, and muscle contraction \citep{Gotti1997}. As an integral membrane protein, the function and organization of nAChR is strongly dictated by its surrounding lipid environment. DHA is an $\omega$-3 polyunsaturated fatty acid abundant in synaptic membranes and the Torpedo electric organs, which are both native nAChR membranes. With its six double bonds, DHA is considered highly disordered and can induce domain formation in membranes \citep{S000930840800032720080101}. We previously observed \citep{Sharp2019} favorable interactions between a single nAChR and DHA-rich, cholesterol-poor domains using coarse-grained simulations. Here, we have extended this approach to investigate the role of lipid topology and domain-formation on boundary lipids of individual nAChRs. We have also conducted the first simulations containing multiple nAChRs, which has allowed us to observe spontaneous dimerization.  

While DHA is implicated in human health and disease, \citep{12439486320170901} experimental studies considering its interactions with nAChR have exclusively considered its free-fatty acid form (FFA), \citep{Antollini2016} 
rather than as an acyl chain component of a phospholipid. Application of $\omega-3$ FFAs causes a significant reduction in open times observed through single-channel recordings  \citep{Bouzat1993}. Here, we observe a substantial effect of lipid topology on both embedded and annular lipids: DHA chains in homoacidic phospholipids are far more likely to be found as either annular or embedded boundary lipids.  We previously observed only quantitative effects of swapping PE with PC on partitioning, but the headgroup does serve to anchor the lipid at the membrane/protein interface  \citep{Sharp2019}.  DHA in its FFA form (without a headgroup) may diffuse into an open nAChR pore, blocking the channel. 

We find that, consistent with coarse-grained MD simulations using one nAChR \citep{Sharp2019}, multiple nAChRs continue to prefer the liquid-disordered phase containing long-chain $\omega$-3 fatty acids. While the number of nAChRs in the system did not affect the partitioning profile in these simulations, it did affect the composition of embedded lipids. Our results are consistent with each nAChR monomer in an oligomer blocking access to embedded lipid binding sites in the other monomers, suggesting an intriguing coupling between specific binding and membrane organization.  

Interestingly, upon removing membrane organization, embedded lipids cluster around specific transmembrane helices in a five-fold symmetry around nAChR. This finding suggests that intrinsic lipid preferences are primarily helix dependent, rather than subunit dependent. In homoacid membranes, a lipid preference for DHA was observed across all transmembrane helices, with shells of PUFAs found even at the border of the liquid-ordered phase. Although saturated acyl chains and cholesterol were generally depleted around nAChR, the highest densities for both lipids were found around the M1 and M3 helices, as seen in heteroacid membranes.

In native membranes, nAChR dimers can be stabilized by a disulfide bond between $\delta$ subunits \citep{Chang1977}.  An early controversy \citep{Anholt1980,Ruechel1981,Zingsheim1982,Schindler1984} concerned whether the disulfide bond was necessary for dimer formation. There is no mechanism for covalent bonds between monomers in these coarse-grained simulations. All dimers were stabilized by non-covalent interactions, consistent with the results of \citep{Ruechel1981,Schindler1984}. We observed far more stable dimers in the homoacid mixtures than the heteroacid mixtures, which would be consistent with high sensitivity to experimental conditions. We did not observe a consistent $\delta-\delta$ preference for interfacing subunits in either homoacid or heteroacid mixtures. Schindler {\it et al} (1984) observed an apparent gain-of-function for single-channels within dimers relative to monomers, regardless of disulfide linking \citep{Schindler1984}, suggesting that lipid modulation of oligomerization could also provide a pathway for modulating single channel function. 

\begin{acknowledgements}
GB was supported by research grants NSF MCB1330728 and NIH P01GM55876-14A1. GB and LM were also supported through a grant from the Research Corporation for Scientific Advancement. This project was supported with computational resources from the National Science Foundation XSEDE program through allocation NSF-MCB110149, a local cluster funded by NSF-DBI1126052, the Rutgers University Office of Advanced Research Computing (OARC) and the Rutgers Discovery Informatics Institute (RDI2), which is supported by Rutgers and the State of New Jersey. We are grateful to Dr. J\'{e}r\^{o}me H\'{e}nin for his helpful suggestions throughout this study.  
\end{acknowledgements}


%
%

\section{Compliance with Ethical Standards}
{\bf Conflict of interest:} The authors declare that they have no conflict of interest. This research was supported in part by the National Science Foundation, the National Institutes of Health, and the Research Corporation for Scientific Advancement. 

\bibliographystyle{spbasic}
\bibliography{BIBx.bib}

\end{document}